%% file: main.tex
\documentclass[12pt,a4paper]{article}

\usepackage{ifthen} 
\newboolean{pdflatex}
\setboolean{pdflatex}{true} 

\newboolean{articletitles}
\setboolean{articletitles}{true} 

\newboolean{uprightparticles}
\setboolean{uprightparticles}{false} 

\newboolean{inbibliography}
\setboolean{inbibliography}{false} 

\input{preamble}
\usepackage{longtable} 

\input{macros}

\begin{document}

\renewcommand{\thefootnote}{\fnsymbol{footnote}}
\setcounter{footnote}{1}

\hypersetup{pageanchor=false}


\input{title}

\renewcommand{\thefootnote}{\arabic{footnote}}
\setcounter{footnote}{0}



\pagestyle{plain} 
\setcounter{page}{1}
\pagenumbering{arabic}

\hypersetup{pageanchor=true}


\input{Introduction}
\input{Isospin}
\input{Resonances}
\input{Missing}
\input{Results}

\setboolean{inbibliography}{true}
\bibliographystyle{LHCb}
\bibliography{main,d0munu,LHCb-PAPER,LHCb-CONF,LHCb-DP,LHCb-TDR}

\newpage
\newpage

\appendix
\input{Appendix}

\end{document}

%% file: preamble.tex
\usepackage[top=1in, bottom=1.25in, left=1in, right=1in]{geometry}

\usepackage{microtype}
\usepackage{lineno}  
\usepackage{xspace} 
\usepackage{caption} 

\usepackage{graphicx}  
\usepackage{color}
\usepackage{colortbl}
\graphicspath{{./figs/}} 

\usepackage{amsmath} 
\usepackage{amssymb}
\usepackage{amsfonts}
\usepackage{upgreek} 

\newcommand*\patchAmsMathEnvironmentForLineno[1]{%
\expandafter\let\csname old#1\expandafter\endcsname\csname #1\endcsname
\expandafter\let\csname oldend#1\expandafter\endcsname\csname
end#1\endcsname
 \renewenvironment{#1}%
   {\linenomath\csname old#1\endcsname}%
   {\csname oldend#1\endcsname\endlinenomath}%
}
\newcommand*\patchBothAmsMathEnvironmentsForLineno[1]{%
  \patchAmsMathEnvironmentForLineno{#1}%
  \patchAmsMathEnvironmentForLineno{#1*}%
}
\AtBeginDocument{%
\patchBothAmsMathEnvironmentsForLineno{equation}%
\patchBothAmsMathEnvironmentsForLineno{align}%
\patchBothAmsMathEnvironmentsForLineno{flalign}%
\patchBothAmsMathEnvironmentsForLineno{alignat}%
\patchBothAmsMathEnvironmentsForLineno{gather}%
\patchBothAmsMathEnvironmentsForLineno{multline}%
\patchBothAmsMathEnvironmentsForLineno{eqnarray}%
}

\usepackage{hyperref}    
\usepackage[all]{hypcap} 

\input{lhcb-symbols-def} 

\usepackage{subcaption}
\usepackage{siunitx}
\DeclareSIUnit\evolt{e\kern -0.1em V}
\usepackage{physics}
\usepackage{booktabs}
\usepackage[capitalize]{cleveref}
\usepackage{multirow}

\usepackage{pdflscape}

\usepackage{cite} 
\usepackage{mciteplus}

%% file: lhcb-symbols-def.tex
\usepackage{xspace} 
\usepackage{upgreek}

\def\MagUp {\mbox{\em Mag\kern -0.05em Up}\xspace}

\ifthenelse{\boolean{uprightparticles}}%
{

 \def\Pnu         {\ensuremath{\upnu}\xspace}                 
                  
 \def\Ppi         {\ensuremath{\uppi}\xspace}

 \def\PDelta      {\ensuremath{\Delta}\xspace}                 
 \def\PXi      {\ensuremath{\Xi}\xspace}                 
 \def\PLambda      {\ensuremath{\Lambda}\xspace}                 
 \def\PSigma      {\ensuremath{\Sigma}\xspace}                 
 \def\POmega      {\ensuremath{\Omega}\xspace}                 
 \def\PUpsilon      {\ensuremath{\Upsilon}\xspace}                 
 
 \def\PB      {\ensuremath{\mathrm{B}}\xspace}                 
                  
 \def\PD      {\ensuremath{\mathrm{D}}\xspace}

 \def\PK      {\ensuremath{\mathrm{K}}\xspace}

 \def\Pi      {\ensuremath{\mathrm{i}}\xspace}

 \def\Ps      {\ensuremath{\mathrm{s}}\xspace}

}
{

 \def\Pnu         {\ensuremath{\nu}\xspace}                 
                  
 \def\Ppi         {\ensuremath{\pi}\xspace}

 \mathchardef\PDelta="7101
 \mathchardef\PXi="7104
 \mathchardef\PLambda="7103
 \mathchardef\PSigma="7106
 \mathchardef\POmega="710A
 \mathchardef\PUpsilon="7107
                  
 \def\PB      {\ensuremath{B}\xspace}                 
                  
 \def\PD      {\ensuremath{D}\xspace}

 \def\PK      {\ensuremath{K}\xspace}

 \def\Pi      {\ensuremath{i}\xspace}

 \def\Ps      {\ensuremath{s}\xspace}

}

\makeatletter
\ifcase \@ptsize \relax
  \newcommand{\miniscule}{\@setfontsize\miniscule{4}{5}}
\or
  \newcommand{\miniscule}{\@setfontsize\miniscule{5}{6}}
\or
  \newcommand{\miniscule}{\@setfontsize\miniscule{5}{6}}
\fi
\makeatother

\DeclareRobustCommand{\optbar}[1]{\shortstack{{\miniscule (\rule[.5ex]{1.25em}{.18mm})}
  \\ [-.7ex] $#1$}}

\def\neub       {{\ensuremath{\overline{\Pnu}}}\xspace}

\def\squark    {{\ensuremath{\Ps}}\xspace}

\def\pion   {{\ensuremath{\Ppi}}\xspace}
\def\piz    {{\ensuremath{\pion^0}}\xspace}

\def\pip    {{\ensuremath{\pion^+}}\xspace}
\def\pim    {{\ensuremath{\pion^-}}\xspace}
\def\pipm   {{\ensuremath{\pion^\pm}}\xspace}

\def\kaon    {{\ensuremath{\PK}}\xspace}
  \def\Kbar    {{\kern 0.2em\overline{\kern -0.2em \PK}{}}\xspace}

\def\KorKbar    {\kern 0.18em\optbar{\kern -0.18em K}{}\xspace}

\def\Km      {{\ensuremath{\kaon^-}}\xspace}

  \def\Dbar    {{\kern 0.2em\overline{\kern -0.2em \PD}{}}\xspace}
\def\D       {{\ensuremath{\PD}}\xspace}

\def\DorDbar    {\kern 0.18em\optbar{\kern -0.18em D}{}\xspace}
\def\Dz      {{\ensuremath{\D^0}}\xspace}

\def\Dp      {{\ensuremath{\D^+}}\xspace}

\def\Dstar   {{\ensuremath{\D^*}}\xspace}

\def\Dstarz  {{\ensuremath{\D^{*0}}}\xspace}

\def\Dstarp  {{\ensuremath{\D^{*+}}}\xspace}

\def\Ds      {{\ensuremath{\D^+_\squark}}\xspace}

\def\Dss     {{\ensuremath{\D^{*+}_\squark}}\xspace}

\def\B       {{\ensuremath{\PB}}\xspace}
\def\Bbar    {{\ensuremath{\kern 0.18em\overline{\kern -0.18em \PB}{}}}\xspace}
\def\Bb      {{\ensuremath{\Bbar}}\xspace}
\def\BorBbar    {\kern 0.18em\optbar{\kern -0.18em B}{}\xspace}

\def\Bu      {{\ensuremath{\B^+}}\xspace}
\def\Bub     {{\ensuremath{\B^-}}\xspace}

\def\Bd      {{\ensuremath{\B^0}}\xspace}

\def\Bdb     {{\ensuremath{\Bbar{}^0}}\xspace}

  \def\Y#1S{\ensuremath{\PUpsilon{(#1S)}}\xspace}

\def\Lbar        {{\ensuremath{\kern 0.1em\overline{\kern -0.1em\PLambda}}}\xspace}
\def\LorLbar    {\kern 0.18em\optbar{\kern -0.18em \PLambda}{}\xspace}

\def\BF         {{\ensuremath{\mathcal{B}}}\xspace}

\newcommand{\decay}[2]{\ensuremath{#1\!\to #2}\xspace}         

\def\to                 {\ensuremath{\rightarrow}\xspace}

\def\AT#1     {\ensuremath{A_{\mathrm{T}}^{#1}}\xspace}           

\def\C#1      {\ensuremath{\mathcal{C}_{#1}}\xspace}                       
\def\Cp#1     {\ensuremath{\mathcal{C}_{#1}^{'}}\xspace}                    
\def\Ceff#1   {\ensuremath{\mathcal{C}_{#1}^{\mathrm{(eff)}}}\xspace}        
\def\Cpeff#1  {\ensuremath{\mathcal{C}_{#1}^{'\mathrm{(eff)}}}\xspace}       
\def\Ope#1    {\ensuremath{\mathcal{O}_{#1}}\xspace}                       
\def\Opep#1   {\ensuremath{\mathcal{O}_{#1}^{'}}\xspace}                    

\newcommand{\tev}{\ifthenelse{\boolean{inbibliography}}{\ensuremath{~T\kern -0.05em eV}\xspace}{\ensuremath{\mathrm{\,Te\kern -0.1em V}}}\xspace}
\newcommand{\gev}{\ensuremath{\mathrm{\,Ge\kern -0.1em V}}\xspace}
\newcommand{\mev}{\ensuremath{\mathrm{\,Me\kern -0.1em V}}\xspace}
\newcommand{\kev}{\ensuremath{\mathrm{\,ke\kern -0.1em V}}\xspace}
\newcommand{\ev}{\ensuremath{\mathrm{\,e\kern -0.1em V}}\xspace}
\newcommand{\gevc}{\ensuremath{{\mathrm{\,Ge\kern -0.1em V\!/}c}}\xspace}
\newcommand{\mevc}{\ensuremath{{\mathrm{\,Me\kern -0.1em V\!/}c}}\xspace}
\newcommand{\gevcc}{\ensuremath{{\mathrm{\,Ge\kern -0.1em V\!/}c^2}}\xspace}
\newcommand{\gevgevcccc}{\ensuremath{{\mathrm{\,Ge\kern -0.1em V^2\!/}c^4}}\xspace}
\newcommand{\mevcc}{\ensuremath{{\mathrm{\,Me\kern -0.1em V\!/}c^2}}\xspace}

\def\gsim{{~\raise.15em\hbox{$>$}\kern-.85em
          \lower.35em\hbox{$\sim$}~}\xspace}
\def\lsim{{~\raise.15em\hbox{$<$}\kern-.85em
          \lower.35em\hbox{$\sim$}~}\xspace}

\def\tell1  {TELL1\xspace}
\def\ukl1   {UKL1\xspace}



%% file: macros.tex
\def\dstst{{\ensuremath{D^{**0}}}\xspace}
\def\dststp{{\ensuremath{D^{**+}}}\xspace}

\def\f12{\ensuremath{\frac{1}{2}}\xspace}

%% file: title.tex
\begin{titlepage}



\vspace*{4.0cm}

{\normalfont\bfseries\boldmath\huge
\begin{center}
 An experimentalist's guide to the semileptonic bottom to charm branching fractions
\end{center}
}

\vspace*{2.0cm}

\begin{center}
M.~Rudolph$^1$.
\bigskip\\
{\normalfont\itshape\footnotesize
$ ^1$Syracuse University, Syracuse, NY, USA
}
\end{center}

\vspace*{1.0cm}
\begin{center}
May 14, 2018
\end{center}

\vspace{\fill}

\begin{abstract}
  \noindent This work summarizes the current status of the measured semileptonic branching fractions \decay{B^{0,+}}{X_c \mu\nu}.  The sum of exclusive measurements is compared with the inclusive determination, accounting for isospin extrapolation. Further derived quantities are computed, taking into account different explanations for the unmeasured components of the total branching fraction. These quantities focus on the charge breakdown of the final states, and are designed for use as inputs or comparisons in future experimental measurements.
\end{abstract}

\vspace*{2.0cm}
\vspace{\fill}

{\footnotesize 
\centerline{Licensed under \href{http://creativecommons.org/licenses/by-nc-sa/4.0/}{CC-BY-NC-SA-4.0}.}}
\vspace*{2mm}

\end{titlepage}

\pagestyle{empty}  


\newpage
\setcounter{page}{2}
\mbox{~}


%% file: Introduction.tex
\section{Introduction}
\label{sec:intro}

The full makeup of the semileptonic width of \Bub and \Bdb mesons\footnote{Charge conjugation is implied throughout this paper} is not fully understood.  Measurements have been made of both the inclusive decay width and a number of decay widths to various exclusive final states.  However, the sum of the measured exclusive decays do not saturate the full inclusive width.  In this paper, we examine the available measurements, extrapolate them to isospin related decays, and provide expectations for derived quantities that may be useful for future experimental comparisons or as inputs to new measurements.

For the determination of the inclusive branching fractions, we use the average from Ref.~\cite{LHCb-PAPER-2016-031}, which relies on measurements of the lepton spectrum from CLEO~\cite{Mahmood:2004kq}, BaBar~\cite{Aubert:2006au}, and Belle~\cite{Urquijo:2006wd}.  
When combining the CLEO measurement with the other two, the average assumes the equality of the  total semileptonic width for \Bu and \Bd, and therefore that the branching fractions differ only by the lifetime difference of \Bu and \Bd. The results of the average are
\begin{align*}
  \BF( \decay{\Bub}{X\mu\nu} ) &= (11.09 \pm 0.20)\% \\
  \BF( \decay{\Bdb}{X\mu\nu} ) &= (10.31 \pm 0.19)\%.
\end{align*}
In the rest of this paper, we consider measurements of \Bu and \Bd branching fractions separately.  We ignore the approximately 1\% contribution of $b\to u$ decays, which is well within the experimental uncertainty.

A number of exclusive branching fraction measurements have been made with final states including $D$, $D^*$, and $D^{(*)}\pi$.  These have been adjusted for updated measurements of their input parameters and then averaged together by the Heavy Flavor Averaging Group (HFLAV)~\cite{Amhis:2016xyh}, and are listed in \cref{tab:meas_bf}. The average of the exclusive \decay{\Bdb}{\Dstarp \ell^-\neub} branching fraction is derived from a global fit using the CLN parameterization~\cite{Caprini:1997mu}.  For the decays \decay{\Bub}{\Dstarz\ell^-\neub} and \decay{\Bb}{\D \ell^-\neub}, we use the one-dimensional averages from HFLAV. Recently, Belle has produced a new measurement for $D^{(*)}\pi$~\cite{Vossen:2018zeg}, which agrees with the previous average but has not yet been included.

\begin{table}[htbp]
  \begin{center}
  \caption{Measured exclusive branching fractions to $D^{(*)}n\pi$ or $D_s^{(*)}K$.  As determined by HFLAV~\cite{Amhis:2016xyh}. The statistical and systematic uncertainties have been totaled.\label{tab:meas_bf}}
  \begin{tabular}{l
      S[table-format = 1.3]@{\,\( \pm \)\,}
      S[table-format = 1.3]}
    \toprule
    Decay & \multicolumn{2}{c}{\BF (\%)} \\
    \midrule
    \decay{\Bub}{\Dz\mu\nu} & 2.33 & 0.1  \\
    \decay{\Bub}{\Dstarz\mu\nu} &  5.59 & 0.19 \\
    \decay{\Bub}{\Dp\pim\mu\nu} & 0.41 & 0.05 \\
    \decay{\Bub}{\Dstarp\pim\mu\nu} & 0.6 &0.06 \\
    \decay{\Bub}{\Ds\Km\mu\nu} & 0.03 & 0.014 \\
    \decay{\Bub}{\Dss\Km\mu\nu} & 0.029 & 0.019 \\
    \midrule
    \decay{\Bdb}{\Dp\mu\nu} & 2.2 & 0.1\\
    \decay{\Bdb}{\Dstarp\mu\nu} & 4.88 & 0.1 \\
    \decay{\Bdb}{\Dz\pip\mu\nu} & 0.42 & 0.06 \\
    \decay{\Bdb}{\Dstarz\pip\mu\nu} & 0.47 & 0.08 \\
    \bottomrule
  \end{tabular}
  \end{center}
\end{table}

Of particular note for this paper is the large variation between different measurements in the \decay{\Bdb}{\Dstarp \ell^- \neub}.  Both the HFLAV and Particle Data Group (PDG)~\cite{Olive:2016xmw} averages for \decay{\Bdb}{\Dstarp \ell^- \neub} are based on global fits that depend on the parameterization, and give a value below 5\%. Some single measurements, such as  the BaBar measurement of the exclusive \Dstar decay~\cite{Aubert:2007qw}
\begin{align*}
  \BF\qty(\decay{\Bdb}{\Dstarp \ell^- \neub}) &= 5.49 \pm  0.16 \text{(stat.)} \pm 0.25 \text{(syst.)} \\
  \BF\qty(\decay{\Bub}{\Dstarz \ell^- \neub}) &= 5.83 \pm  0.15 \text{(stat.)} \pm 0.30 \text{(syst.)}
\end{align*}
are larger. This measurement is based on yields extracted from missing mass distributions in fully reconstructed tagged decays.

The HFLAV averaging procedure takes into account correlations between the measurements when producing an average, but do not quote correlations between different averages that come from related measurements.  In particular, the $D\pi$ exclusive rates come from a pair of measurements by BaBar and Belle.  BaBar determines them normalized to the total semileptonic branching fraction, while Belle uses the exclusive decay to $D\ell\nu$. In our derived quantities, we will take the total uncertainty on these measurements as 100\% correlated with the exclusive $\D\ell\nu$ branching fractions.   The $\Dstar\ell\nu$ states will be considered uncorrelated.

Final states with  two charged pions have been measured as ratios over the corresponding no pion exclusive states~\cite{Lees:2015eya}:
\begin{align*}
  \BF\qty(\decay{\Bbar}{\D\pip\pim\mu\nu})/\BF\qty(\decay{\Bbar}{\D\mu\nu}) &= \qty( 6.7 \pm 1.3 )\% \\
  \BF\qty(\decay{\Bbar}{\Dstar\pip\pim\mu\nu})/\BF\qty(\decay{\Bbar}{\Dstar\mu\nu}) &= \qty( 1.9 \pm 0.6 )\%
\end{align*}

Exclusive decays to particular excited resonances have also been measured, and will be discussed in more detail in \cref{sec:resonances}.

To determine the production of final state charm mesons from more excited decays, the following branching fractions from the PDG are also used~\cite{Olive:2016xmw}:
\begin{align*}
  \BF\qty(\decay{\Dstarz}{\Dz X}) &= 100\% \\
  \BF\qty(\decay{\Dstarp}{\Dz X}) &= \qty( 67.7 \pm 0.5 )\%  
\end{align*}

Other strategies have previously been employed to study the discrepancy between inclusive and exclusive measurements.  A more detailed analysis of the leptonic and hadronic moments~\cite{Bernlochner:2014dca} beyond the extraction of the total width has been conducted. It is difficult to draw definitive conclusions, but in general it does not suggest a resolution of the inclusive--exclusive gap involving large contributions from higher excited states. It does suggest that a higher value for the exclusive \Dstar branching fraction is probable.

In this paper, we will first discuss the extrapolation of the measured branching fractions to isospin related channels in \cref{sec:iso}, and the resulting total inclusive--exclusive gap in \cref{sec:missing}.  We will conclude by deriving a number of related quantities that may be useful for future measurements, and discuss the possible applications in \cref{sec:results,sec:conclusion}.

%% file: Isospin.tex
\section{Extrapolation for \texorpdfstring{$D\pi$}{Dpi} and \texorpdfstring{$D2\pi$}{D2pi}}
\label{sec:iso}

The measurements of $D$ or \Dstar production with one or two additional pions only include final states with charged pions.  We therefore need to extrapolate to the corresponding decays with neutral pions to calculate the total rates.  In the one pion case this can be done straightforwardly using isospin.  The result is
\begin{equation*}
\frac{\BF\qty( \decay{\Bb}{D^{(*)} \pipm \ell\nu})}{ \BF\qty( \decay{\Bb}{D^{(*)} \piz \ell\nu})} = 2
\end{equation*}

One potential issue is the separation between \Dstarp and $\Dz\pip$ non-resonant states~\cite{PhysRevD.48.3204} near the \Dstarp mass.  In this paper, we consider the experimental measurements of \Dstar to include the full resonant contribution so that isospin can be applied.  

The proper extrapolation to all final states with two additional pions is less clear.  We consider multiple possibilities:
\begin{description}
\item[Sequential emission of two single pions.]  The isospin argument is the same as the single pion case applied twice.  Assigning $2/3$ probability to charged pion and $1/3$ for neutral, there are nine combinations possible; 4 $\pip\pim$, 4$\pi^\pm \piz$, 1 $\piz\piz$.
\item[Di-pion state with definite isospin.]  If the di-pion system has a definite isospin (1 or 0), then one can make the calculations for isospin by first combining the two pions.  For $I=1$, \pip\pim makes up one third of the rate, with the rest $\pip\piz$.  For $I=0$, two thirds is \pip\pim with one third \piz\piz.
\item[Other isospin combination.]  Both di-pion isospin states could be allowed, potentially with different amplitudes.  This can have a large effect -- for $\pip\pim$ the two potential di-pion states enter the amplitude with opposite signs; however, in this work we do not consider these models for the extrapolation.
\end{description}
In all likelihood, the total two pion rate is a mixture of these possibilities.  Some of the rate is certainly from measured resonance decays, such as \decay{D_1(2420)}{D \pip\pim}~\cite{Abe:2004sm,LHCb-PAPER-2011-016}.  Radially excited $D$ and \Dstar mesons can decay via two pion modes~\cite{Bernlochner:2012bc} as well as two single pion emissions through the $L=1$ resonances.

For the purposes of deriving the total exclusive rate we choose
\begin{equation*}
  \frac{\BF\qty( \decay{\Bb}{D^{(*)} \pip\pim \ell\nu})}{\BF\qty( \decay{\Bb}{D^{(*)} \pi\pi \ell\nu})} = 0.5 \pm 0.17.
\end{equation*}
This same choice was used by BaBar~\cite{Lees:2015eya} and supported by Ref.~\cite{Bernlochner:2016bci}.

In order to derive expectations for final states involving \Dz and \Dp separately, a more specific choice must be made, since these depend on the fractions of $\pipm\piz$ and $\piz\piz$ in the unmeasured part.  We have chosen sequential emission for the central values we quote.  We also perform alternate calculations assuming di-pion $I=0$ and $I=1$.  

After performing the isospin extrapolation, we derive the rates for \Bu and \Bd to decay to \Dz and \Dp final states with up to two additional pions not from a \Dstar decay.  The results are listed in \cref{tab:finalstates}.

\begin{table}[htb]
    \caption{Summary of measured and isospin related branching fractions of \Bu and \Bd to final states with \Dz or \Dp.  States with different numbers of initial pions are separated.  The top part of the table is filled assuming sequential emission for the two pion case.  The total fractions instead assuming di-pion $I=0$ or $I=1$ are listed below.\label{tab:finalstates}}
  \begin{center}
\begin{tabular}{l
    S[table-format = 1.3]@{\,\( \pm \)\,}
    S[table-format = 1.3]
    S[table-format = 1.3]@{\,\( \pm \)\,}
    S[table-format = 1.3]
    S[table-format = 1.3]@{\,\( \pm \)\,}
    S[table-format = 1.3]
    S[table-format = 1.3]@{\,\( \pm \)\,}
    S[table-format = 1.3]}
\toprule
 & \multicolumn{4}{c}{\Bub ~\BF (\%)} & \multicolumn{4}{c}{\Bdb ~\BF (\%)} \\
 Initial state & \multicolumn{2}{c}{\Dz} &  \multicolumn{2}{c}{\Dp} &  \multicolumn{2}{c}{\Dz} &  \multicolumn{2}{c}{\Dp} \\
\midrule
\D           $\mu\nu$ & 2.33 & 0.10 & \multicolumn{2}{c}{---} & \multicolumn{2}{c}{---} & 2.20 & 0.10 \\
\Dstar       $\mu\nu$ & 5.59 & 0.19 & \multicolumn{2}{c}{---} & 3.30 & 0.07 & 1.58 & 0.04 \\
\D\pion      $\mu\nu$ & 0.205 & 0.025 & 0.41 & 0.05 & 0.42 & 0.05 & 0.210 & 0.025 \\
\Dstar\pion  $\mu\nu$ & 0.71 & 0.07 & 0.194 & 0.020 & 0.63 & 0.08 & 0.076 & 0.010 \\
\D 2\pion    $\mu\nu$ & 0.20 & 0.04 & 0.156 & 0.031 & 0.147 & 0.029 & 0.18 & 0.04 \\
\Dstar 2\pion$\mu\nu$ & 0.20 & 0.07 & 0.034 & 0.011 & 0.17 & 0.05 & 0.037 & 0.012 \\
\midrule
Total         & 9.23 & 0.29 & 0.79 & 0.08 & 4.67 & 0.17 & 4.28 & 0.15 \\
\midrule
$I=0$ total & 9.22 & 0.29 & 0.60 & 0.07 & 4.45 & 0.15 & 4.33 & 0.16 \\
$I=1$ total & 9.24 & 0.30 & 0.98 & 0.10 & 4.90 & 0.19 & 4.24 & 0.15 \\
\bottomrule
\end{tabular}
  \end{center}
\end{table}



%% file: Resonances.tex
\section{Resonant contributions}
\label{sec:resonances}

Measurements of the resonant contribution to the $D^{(*)}\pi$ states have also been carried out.  These contributions come from the four \D mesons with orbital angular momentum $L=1$, the relatively narrow  $D_1(2420)$ and  $D_2^*(2460)$ states and the wider $D_0^*(2400)$ and $D_1'(2430)$.  HFLAV quotes four averaged branching fractions to either $D\pi$ or $\Dstar\pi$ final states; the last allowed decay \decay{D_2^*}{D\pi} is not averaged.  To account for this mode, we use the measurement from BaBar~\cite{Aubert:2008zc}, which quotes the fraction of all $D_2^*$ decays in $D\pi$
\begin{equation*}
   \frac{\BF\qty(\decay{D_2^*}{D\pim})}{\BF\qty(\decay{D_2^*}{D^{(*)}\pim})} = 0.62 \pm 0.03 \pm 0.02.
\end{equation*}
A summary of the results for \Bub is found in \cref{tab:resdstst}.


\begin{table}[htbp]
  \caption{HFLAV averages for resonant $\decay{\Bub}{D^{**}\ell\nu}$ production.  The exception is \decay{\Bub}{\decay{D_2^*}{D^{*+}\pim}\ell\nu}  which is only from BaBar's measurement of \decay{\Bub}{\decay{D_2^{*0}}{D^{(*)+}\ell\nu}} and $\BF_{D/D^{(*)}}$ and for which only the total uncertainty is quoted.\label{tab:resdstst}}
  \begin{center}
    \begin{tabular}{l
      S[table-format = 1.3]@{\,\( \pm \)\,}
      S[table-format = 1.3]@{\,\( \pm \)\,}
      S[table-format = 1.3]
      }
      \toprule
      Decay & \multicolumn{3}{c}{ \BF (\%)${}\pm \text{stat.} \pm \text{syst.}$  } \\
      \midrule
      \decay{\Bub}{\decay{D_1}{D^{*+}\pim}\ell\nu} & 0.281 & 0.010 & 0.015 \\
      \decay{\Bub}{\decay{D_1'}{D^{*+}\pim}\ell\nu} & 0.13 & 0.03 & 0.02 \\
      \decay{\Bub}{\decay{D_0^*}{D^{+}\pim}\ell\nu} & 0.28 & 0.03 & 0.04 \\
      \decay{\Bub}{\decay{D_2^*}{D^{*+}\pim}\ell\nu} & 0.077 & 0.006 & 0.004 \\
      \midrule
      \decay{\Bub}{\decay{D_2^*}{D^{+}\pim}\ell\nu} & 0.142 & \multicolumn{2}{S[table-format=1.3]}{ 0.021} \\
      \bottomrule
    \end{tabular}
  \end{center}
\end{table}

The HFLAV averages do not include \Bd decays.  The BaBar measurement referenced above~\cite{Aubert:2008zc} also quotes results for \decay{\Bdb}{D_2^{*+}\ell\nu} and \decay{\Bdb}{D_1^+\ell\nu},
\begin{align*}
   \BF\qty(\decay{\Bdb}{D_1^+\ell\nu})\BF\qty(\decay{D_1^+}{D^{*0}\pip}) &= \qty( 2.78 \pm 0.24 \pm 0.25 )\times 10^{-3} \\
   \BF\qty(\decay{\Bdb}{D_2^{*+}\ell\nu})\BF\qty(\decay{D_2^{*+}}{D^{(*)0}\pip}) &= \qty( 1.77 \pm 0.26 \pm 0.11 )\times 10^{-3} 
\end{align*}
and a separate paper~\cite{Aubert:2008ea} quotes
\begin{align*}
  \BF\qty(\decay{\Bdb}{D_1^+\ell\nu})\BF\qty(\decay{D_1^+}{\Dstarz \pip}) &= (0.27 \pm 0.04 \pm 0.03)\% \\
  \BF\qty(\decay{\Bdb}{D_2^{*+}\ell\nu})\BF\qty(\decay{D_2^{*+}}{\Dz \pip}) &= (0.07 \pm 0.03 \pm 0.01)\% \\
  \BF\qty(\decay{\Bdb}{D_1^{\prime +}\ell\nu})\BF\qty(\decay{D_1^{\prime +}}{\Dstarz \pip}) & = (0.31 \pm 0.07 \pm 0.05)\% \\
  \BF\qty(\decay{\Bdb}{D_0^{*+}\ell\nu})\BF\qty(\decay{D_0^{*+}}{\Dz\pip}) &= (0.44 \pm 0.08 \pm 0.06)\%,
\end{align*}
where the $D_2^{*+}$ channel only includes the decay to $D\pi$.  The two results are in agreement with one another.  To compute the sums we take the narrow state results from Ref.~\cite{Aubert:2008zc}, and the wide states from Ref.~\cite{Aubert:2008ea}.  

When computing the sums, we take the measurements to be fully correlated.  The total results for the four considered channels are
\begin{align*}
\decay{ \Bub }{ \Dp\pim } &= (0.42 \pm 0.07)\% \\
\decay{ \Bub }{ \Dstarp\pim } &= (0.50 \pm 0.07)\% \\
\decay{ \Bdb }{ \Dz\pip } &= (0.55 \pm 0.12)\% \\
\decay{ \Bdb }{ \Dstarz\pip } &= (0.41 \pm 0.13)\%
\end{align*}
These numbers are consistent within uncertainties with $D^{(*)}\pi$ production being dominated by resonances.

In addition to these measured rates with one pion, there is also an allowed decay mode \decay{\D_1}{\D \pi \pi}.  An extrapolation to this contribution can be found by applying the measured ratio in the \decay{B}{D_1 \pi} final state\cite{Aaij:2011rj} where both have been observed.  The result from Ref.~\cite{Bernlochner:2016bci} is
\begin{equation*}
  f_{D_1} = \frac{ \BF\qty( \decay{\D_1^0}{ \Dstarp \pim}) }{ \BF\qty( \decay{\D_1^0}{  \Dz \pip\pim})} = 2.32 \pm 0.54.
\end{equation*}
Combined with an isospin extrapolation to the neutral pion mode, the observed \decay{ B}{D\pi\pi} rate may also be dominated by resonance decays.  This argument does not apply to the $\Dstar\pi\pi$ final state. The rate of $\Dstar\pi\pi$ relative to exclusive \Dstar is, however, measured to be much lower than $D\pi\pi$.

%% file: Missing.tex
\section{Extrapolation to the missing semileptonic fraction}
\label{sec:missing}

After adding up the measured branching fractions, and including the extrapolation to final states with neutral pions, we obtain the following missing branching fractions:
\begin{align*}
\Bub \;\BF \text{ missing}  &= (1.0 \pm 0.4 \pm 0.2)\% \\ 
\Bdb \;\BF \text{ missing}  &= (1.3 \pm 0.4 \pm 0.2)\%. 
\end{align*}
The first uncertainty listed comes from the measurements.  The second uncertainty accounts for the neutral pion extrapolation.  In particular, the larger missing fraction is obtained assuming two pions are emitted in an $I=0$ state, while the smaller fraction is obtained for the $I=1$ state.
If only the BaBar measurements for $\BF\qty(\decay{\Bb}{\Dstar \ell^- \neub})$~\cite{Aubert:2007qw} are used, this reduces the missing portions, yielding rates of $\qty(0.7 \pm 0.5)\%$ for \Bdb and $\qty(0.8 \pm 0.5)\%$ for \Bub.

Even considering the large uncertainties, this sum of exclusive branching fractions is in tension with the expectation for the total rate.  To account for this in our further derived results, we use different extrapolation assumptions, and re-calculate the results for each one.  The full envelope of the resulting set will be given.  Depending on the use case for these results, this envelope may or may not be treated as a systematic uncertainty.  A measurement may instead by attempting to distinguish between two of the considered scenarios.  In our main results we will quote the highest and lowest variations for a particular quantity.  Full results for each considered extrapolation strategy  are found in \cref{app:fullresults}.

The following extrapolation strategies are used:
\begin{description}
\item[No extrapolation --] the measured exclusive rates are taken as is.  This strategy assumes that the discrepancy is due to a correlated uncertainty, either in each measured exclusive rate, or in the inclusive rate, and that the contribution from more than two pion production is negligible.
\item[As two pion --] the total missing rate is divided up into \Dz and \Dp final states based on the fractions expected from two pion production.  This strategy assumes that either the two pion rate is under-measured, or that higher pion states dominate the excess and have a similar charge ratio.
\item[Half and half --] the total missing rate is divided equally between \Dz and \Dp final states.  This assumes that the missing rate is largely due to states producing more than two additional pions.  Using the sequential emission strategy discussed in \cref{sec:iso}, the fraction of \Dz produced quickly nears 50\% for the emission of three or more pions.
\item[High \Dstar --] the missing fraction is smaller because higher values for the \Dstar branching fractions are used. The remaining extrapolation is performed as in the two pion case.
\end{description}
Extrapolation as two pion, using sequential emission for the neutral modes, will be quoted as the central values of the envelopes.  This is done primarily for expedience -- this assumption is found to lie in the middle of the extrapolation envelope for most derived quantities.

%% file: Results.tex
\section{Results}
\label{sec:results}

Based on the measurements and extrapolations described previously, we derive results for a number of different quantities that may be useful as observables or as constraints for other measurements. Most of these derived quantities will depend on the relationship between \Dz and \Dp final states, giving them sensitivity to the makeup of the missing rate without fully reconstructing all particles in the decay.

We first summarize the \Bu and \Bd contributions to the \Dz and \Dp final states, for measurements in which the two initial states are not separated. Assuming equal production of \Bu and \Bd, the total ratio of $\Dp X$ final states to $\Dz X$ states, or the fraction of $\Dp X$, is
\begin{align*}
  R_{+/0}         &= 0.387 \pm 0.012 \pm 0.026 \\ 
  f_{+}           &= 0.279 \pm 0.006 \pm 0.014,  
\end{align*}
where the first variation corresponds to the total experimental uncertainty, and the second to the full envelope of all isospin and missing fraction extrapolations.  The main uncertainty in the extrapolation envelope comes from the treatment of the missing fraction; no extrapolation provides lower values and half-and-half extrapolation the higher values. Each individual result used to create the envelope, for this and the following results, is found in \cref{app:fullresults}.

We can split this charge ratio up further if the initial \B flavor can also be determined. The results
\begin{align*}
  R_{+/0, \Bub} &= 0.113 \pm 0.010 ^{+0.032}_{-0.057} \\
  R_{+/0,\Bdb} &= 0.89 \pm 0.03 ^{+0.24}_{-0.10} \\
\end{align*}
are particularly sensitive to the extrapolation to the unmeasured decays.

Next, we instead examine the fractional contribution of \Bdb decays to either charge final state,
\begin{align*}
  f_{\Bdb,\Dz}     &= 0.354 \pm 0.006 ^{+0.019}_{ -0.039 } \\ 
  f_{\Bdb,\Dp}     &= 0.812 \pm 0.014 ^{+0.088}_{ -0.045 }.  
\end{align*}
The most important factors in the extrapolation for these quantities is the isospin extrapolation of the two pion case, and then its use in missing fraction.  Requiring di-pion $I=1$ increases the amount of \Dz produced from \Bdb states, while $I=0$ decreases it.

A related set of quantities break this contribution down further by relating how often charged and neutral $D^*$ and $D^{**}$ states produce \Dz final states.  These are quoted as the ratio of \Bdb to \Bub production,
\begin{align*}
r_\Dstar &= \BF\qty(\decay{\Dstarp}{\Dz X}) \frac{ \BF\qty( \decay{\Bdb}{\Dstarp\mu\nu}) }{ \BF\qty( \decay{\Bub}{\Dstarz\mu\nu}) }      & &= 0.591 \pm 0.024 \\ 
r_{D^{**}} &=\frac{ \BF\qty( \decay{D^{**+}}{\Dz X} ) }{ \BF\qty( \decay{D^{**0}}{\Dz X} ) } \frac{ \BF\qty( \decay{\Bdb}{D^{**+}\mu\nu}) }{ \BF\qty( \decay{\Bub}{D^{**0}\mu\nu}) }      & &= 1.07 \pm 0.13 ^{+0.33}_{ -0.46 }.  
\end{align*}
The uncertainty on the first quantity is dominated by the exclusive \Dstar branching fractions.  The largest experimental uncertainties on the second quantity are the exclusive rates to \Dstar with one additional pion.  The extrapolation envelope is also dominated by the isospin extrapolation of the two pion case to the missing fraction.

If the different \B initial states can be separated, as in $b$-factory measurements with a fully reconstructed opposite side decay, we derive expected fractions dividing the exclusive \D and \Dstar from higher excited states. The extrapolation here takes a simpler form since the uncertainty in the missing portion is entirely within the $D^{**}$ fraction.  The nominal value includes the missing fraction; the second uncertainty lists the change when removing it.  For \Bu the results are 
\begin{align*}
  f_{\Dz} &= \BF\qty( \Bub \to \D\mu\nu )/\BF\qty( \Bub \to \D X \mu\nu )       &   &= 0.211 \pm 0.010 +0.026 \\ 
f_{\Dstarz} &= \BF\qty( \Bub \to \Dstar\mu\nu )/\BF\qty( \Bub \to \D X \mu\nu )  &    &= 0.507 \pm 0.020 +0.062 \\ 
f_{\dstst}  &= \BF\qty( \Bub \to D^{**}\mu\nu )/\BF\qty( \Bub \to \D X \mu\nu )   &   &= 0.282 \pm 0.023  -0.088.  
\end{align*}
For \Bd the results are
\begin{align*}
f_{\Dp}    &= \BF\qty( \Bdb \to \D\mu\nu )/\BF\qty( \Bdb \to \D X \mu\nu )       &  &= 0.215 \pm 0.011 +0.033  \\ 
f_{\Dstarp}&= \BF\qty( \Bdb \to \Dstar\mu\nu )/\BF\qty( \Bdb \to \D X \mu\nu )    &  &= 0.476 \pm 0.013 +0.080  \\ 
f_{D^{**+}}   &= \BF\qty( \Bdb \to D^{**}\mu\nu )/\BF\qty(  \Bdb \to\D X \mu\nu )  &  &= 0.310 \pm 0.019 -0.116  
\end{align*}
These fractions correspond to the measurements made by BaBar~\cite{Aubert:2007bq}; we expect this measurement to have a large overlap with the BaBar results from Ref.~\cite{Aubert:2007qw} which utilize a similar technique.  Both results favor a smaller $D^{**}$ component, corresponding more closely to the no extrapolation case.

We also derive similar expectations, restricting the final state to a particular $D$ charge (\decay{\Bub}{\Dz X\ell\nu} and \decay{\Bdb}{\Dp X\ell\nu}).  The results for \Bub are:
\begin{align*}
f_{\Dz,\Dz X} &= \BF\qty( \Bub \to \Dz\mu\nu )/\BF\qty(  \Bub \to\Dz X \mu\nu )     &  &= 0.235 \pm 0.011 ^{+0.018}_{ -0.012 } \\ 
f_{\Dstarz,\Dz X} &= \BF\qty( \Bub \to \Dstarz\mu\nu )/\BF\qty(  \Bub \to\Dz X \mu\nu ) &    &= 0.564 \pm 0.017 ^{+0.042}_{ -0.028 } \\ 
f_{\dstst,\Dz X}  &= \BF\qty( \Bub \to \dstst\mu\nu )/\BF\qty(  \Bub \to\Dz X \mu\nu ) &   &= 0.201 \pm 0.020 ^{+0.039}_{ -0.06 }.  
\end{align*}
The corresponding results for \Bdb are:
\begin{align*}
f_{\Dp,\Dp X}  &= \BF\qty( \Bdb \to \Dp\mu\nu )/\BF\qty(  \Bdb \to\Dp X \mu\nu )       &    &= 0.457 \pm 0.020 ^{+0.062}_{ -0.051 } \\ 
f_{\Dstarp,\Dp X} &= \BF\qty( \Bdb \to \Dstarp\mu\nu )/\BF\qty(  \Bdb \to\Dp X \mu\nu ) &    &= 0.327 \pm 0.011 ^{+0.046}_{ -0.036 } \\ 
f_{\dststp,\Dp X}   &= \BF\qty( \Bdb \to \dststp\mu\nu )/\BF\qty(  \Bdb \to\Dp X \mu\nu ) &  &= 0.216 \pm 0.024 ^{+0.087}_{ -0.107 }.  
\end{align*}
In both cases the extrapolation uncertainty comes from the treatment of the two pion case and the missing fraction, and dominates over the other experimental uncertainties.

\section{Conclusions}
\label{sec:conclusion}

Further experimental effort is clearly required to have a more firm picture of the makeup of the \B semileptonic rate to different charm states.  To address the missing part of the inclusive rate, it is necessary to improve our understanding of the exclusive rate to \Dstar, which has a large discrepancy between different measurements.  Additional measurements of the production of states with two or more emitted pions will also be needed, and may be possible with large data sets at LHCb or Belle II.

For the measurements with additional pions, there are so far only results for the charged final states.  While isospin symmetry provides a strong constraint on the emission of a single neutral pion, the situation is less clear for two or more additional pions.  One possibility would be additional measurements, perhaps of $D^{(*)}\pip\piz$ final states. If precision measurements can be made separating out the production of final states with different \D meson charges from each initial \B flavor, this may also help to constrain this extrapolation.

For example, measurements of inclusive rates to differently charged final states, \emph{i.e.} \decay{\Bub}{\Dz X\mu\nu} compared to \decay{\Bub}{\Dp X \mu\nu}, may be helpful to constrain the contribution from unmeasured decays with two or more pions. Doing so requires tagging the production of the \Bub without reference to the final state, however.

For most of the derived quantities considered in this work,  generally the extrapolation strategy used causes an uncertainty larger than the experimental inputs. This happens because they relate the \Bub and \Bdb decays to \Dz and \Dp without reference to their overall normalizations.  Experimental measurements focused on these relative quantities provide another handle on the missing rate beyond simply measuring new exclusive final states.

Measurements of the relative branching fractions between \D, \Dstar, and $D^{**}$ are particularly sensitive since the extrapolation uncertainty is predominant in the $D^{**}$ portion.  The BaBar measurement~\cite{Aubert:2007bq} already favors a scenario in which the missing fraction is not made up entirely of $D^{**}$ decays. Future precise determinations of these rates may provide more clarity.

At LHCb it is more difficult to separate contributions from \Bub and \Bdb, however it may be possible to extract some information on the fraction of more excited states.  Looking at relative quantities such as the ratio of total \Dp to \Dz production should also be possible.  Combined measurements in different final state channels may also be helpful using the large data sets available.

In summary, we have discussed the current state of the exclusive branching fraction measurements for \decay{\Bb}{X_c\ell\nu}. We then derived expectations for a number of related quantities. Future measurements of these will provide a complementary approach to further direct refinements of exclusive branching fractions.  This knowledge will help determine if the missing semileptonic rate is due to unmeasured excited states, or some other systematic effect.

\section{Acknowledgments}

I would like to thank my colleagues at Syracuse University and on the LHCb collaboration for the discussions that lead to the results presented here. I acknowledge the support of the National Science Foundation, under grant number 1606458, while undertaking this work.

%% file: Appendix.tex
\begin{landscape}

\section{Detailed results}
\label{app:fullresults}

  \begin{table}[htbp]
    \begin{center}
  \caption{Full results for derived quantities for each extrapolation option used in the evelope. The different extrapolations for isospin and the missing rate are described in the text. No ext. refers to scenarios without extrapolation to the missing rate, but the extrapolation to neutral pion states is included.\label{tab:fullresults1}}
  \begin{tabular}{l
    S[table-format = 1.3]
    S[table-format = 1.3]
    S[table-format = 1.3]
    S[table-format = 1.3]
    S[table-format = 1.3]
    S[table-format = 1.3]
    S[table-format = 1.3]
    S[table-format = 1.3]
    S[table-format = 1.3]
    S[table-format = 1.3]
}
    \toprule
    & $\Bu \text{ miss}$ & $\Bd \text{ miss}$ & $R_{+/0}$ & $f_{+}$ & $R_{\Bu}$ & $R_{\Bd}$ & $f_{\Bd,\Dz}$ & $f_{\Bd,\Dp}$ & $r_{\Dstar}$ & $r_{D^{**}}$ \\
\midrule
All $2\pi$         & 1.065 & 1.355 & 0.388 & 0.279 & 0.114 & 0.885 & 0.355 & 0.810 & 0.591 & 1.066 \\
No ext.            & 1.065 & 1.355 & 0.365 & 0.268 & 0.086 & 0.917 & 0.336 & 0.844 & 0.591 & 1.043 \\
50/50              & 1.065 & 1.355 & 0.410 & 0.291 & 0.136 & 0.926 & 0.348 & 0.784 & 0.591 & 1.031 \\
High \Dstar        & 0.815 & 0.719 & 0.374 & 0.272 & 0.105 & 0.862 & 0.356 & 0.819 & 0.638 & 0.971 \\
$I=0$ no ext.      & 1.262 & 1.535 & 0.361 & 0.265 & 0.065 & 0.973 & 0.325 & 0.878 & 0.591 & 0.876 \\
$I=1$ no ext.      & 0.868 & 1.175 & 0.370 & 0.270 & 0.107 & 0.866 & 0.346 & 0.812 & 0.591 & 1.209 \\
$I=0$ all $2\pi$   & 1.262 & 1.535 & 0.396 & 0.283 & 0.058 & 1.127 & 0.316 & 0.900 & 0.591 & 0.602 \\
$I=1$ all $2\pi$   & 0.868 & 1.175 & 0.384 & 0.277 & 0.145 & 0.783 & 0.374 & 0.763 & 0.591 & 1.403 \\
$I=0$ 50/50        & 1.262 & 1.535 & 0.415 & 0.293 & 0.125 & 0.977 & 0.340 & 0.801 & 0.591 & 0.917 \\
$I=1$ 50/50        & 0.868 & 1.175 & 0.406 & 0.289 & 0.147 & 0.877 & 0.355 & 0.767 & 0.591 & 1.157 \\
\bottomrule
\end{tabular}
    \end{center}
  \end{table}

  \begin{table}[htbp]
    \begin{center}
  \caption{Full results for derived quantities for each extrapolation option used in the evelope. The different extrapolations for isospin and the missing rate are described in the text. No ext. refers to scenarios without extrapolation to the missing rate, but the extrapolation to neutral pion states is included. \label{tab:fullresults2}}
  \begin{tabular}{l
    S[table-format = 1.3]
    S[table-format = 1.3]
    S[table-format = 1.3]
    S[table-format = 1.3]
    S[table-format = 1.3]
    S[table-format = 1.3]
    S[table-format = 1.3]
    S[table-format = 1.3]
    S[table-format = 1.3]
    S[table-format = 1.3]
    S[table-format = 1.3]
    S[table-format = 1.3]
    }
    \toprule
    & $f_{\Dz}$ & $f_{\Dstarz}$ & $f_{\dstst}$ & $f_{\Dp}$ & $f_{\Dstarp}$ & $f_{D^{**+}}$ & $f_{\Dz,\Dz X}$ & $f_{\Dstarz,\Dz X}$ & $f_{\dstst,\Dz X}$ & $f_{\Dp,\Dp X}$ & $f_{\Dstarp,\Dp X}$ & $f_{D^{**+},\Dp X}$ \\
    \midrule
All $2\pi$          & 0.210 & 0.504 & 0.286 & 0.213 & 0.473 & 0.313 & 0.234 & 0.562 & 0.204 & 0.455 & 0.326 & 0.220 \\
No ext.             & 0.232 & 0.558 & 0.210 & 0.246 & 0.545 & 0.209 & 0.252 & 0.606 & 0.142 & 0.514 & 0.368 & 0.118 \\
50/50               & 0.210 & 0.504 & 0.286 & 0.220 & 0.487 & 0.293 & 0.239 & 0.573 & 0.189 & 0.457 & 0.327 & 0.216 \\
High \Dstar         & 0.210 & 0.526 & 0.264 & 0.213 & 0.532 & 0.254 & 0.232 & 0.581 & 0.187 & 0.461 & 0.372 & 0.168 \\
$I=0$ no ext.       & 0.237 & 0.569 & 0.194 & 0.251 & 0.556 & 0.193 & 0.253 & 0.606 & 0.141 & 0.508 & 0.364 & 0.128 \\
$I=1$ no ext.       & 0.228 & 0.547 & 0.225 & 0.241 & 0.534 & 0.225 & 0.252 & 0.605 & 0.143 & 0.519 & 0.372 & 0.109 \\
$I=0$ all $2\pi$    & 0.210 & 0.504 & 0.286 & 0.213 & 0.473 & 0.313 & 0.222 & 0.533 & 0.245 & 0.403 & 0.289 & 0.309 \\
$I=1$ all $2\pi$    & 0.210 & 0.504 & 0.286 & 0.213 & 0.473 & 0.313 & 0.241 & 0.577 & 0.182 & 0.486 & 0.348 & 0.166 \\
$I=0$ 50/50         & 0.210 & 0.504 & 0.286 & 0.219 & 0.486 & 0.295 & 0.236 & 0.567 & 0.196 & 0.444 & 0.318 & 0.238 \\
$I=1$ 50/50         & 0.210 & 0.504 & 0.286 & 0.220 & 0.488 & 0.292 & 0.241 & 0.578 & 0.181 & 0.471 & 0.337 & 0.192 \\
\bottomrule
  \end{tabular}
    \end{center}
  \end{table}

\end{landscape}